\documentclass[preprint,aps,showpacs,nofootinbib,preprintnumbers,amsmath,amssymb]{revtex4-1}
\usepackage{}
\usepackage{epsfig}
\usepackage{subfigure}
\usepackage{dcolumn}
\usepackage{bm}
\usepackage[usenames ,dvipsnames]{xcolor}
\usepackage{slashed}
\usepackage{graphicx,color}
\usepackage{epstopdf}
\begin{document}
\title{Roles of scalar mesons in charmless $\Lambda_b$ decays} 

\author{Y.~K. Hsiao$^{1,2,3}$, Yu-Heng Lin$^{3}$, Yao Yu$^{1}$, and C.~Q. Geng$^{1,2,3}$}
\affiliation{
$^1$Chongqing University of Posts \& Telecommunications, Chongqing, 400065, China\\
$^2$Physics Division, National Center for Theoretical Sciences, Hsinchu, Taiwan 300\\
$^3$Department of Physics, National Tsing Hua University, Hsinchu, Taiwan 300
}
\date{\today}

\begin{abstract}
We first study the charmless two-body $\Lambda_b$ decays 
with the scalar mesons as the final states and 
predict that ${\cal B}(\Lambda_b\to \Lambda f_0(980,1500))=
(2.9\pm 0.7,12.4\pm 3.8)\times 10^{-6}$ and 
${\cal B}(\Lambda_b\to p K^{*-}_0(800,1430))=(1.9\pm 0.5,14.1\pm 4.5)\times 10^{-6}$.
With the resonant $f_0(980,1500)\to (\pi^+\pi^-,K^+K^-)$ and $K^{*-}_0\to \bar K^0 \pi^-$
decays, we then obtain
${\cal B}(\Lambda_b\to \Lambda (\pi^+\pi^-,K^+K^-))=(4.2\pm 1.0,3.5\pm0.7)\times 10^{-6}$
and ${\cal B}(\Lambda_b\to p\bar K^0 \pi^-)=(10.4\pm2.9)\times 10^{-6}$, in comparison with
the data  of $(4.6\pm 1.9,15.9\pm2.6)\times 10^{-6}$ and $(12.6\pm 4.0)\times 10^{-6}$ from LHCb,
respectively. 
Our results for $\Lambda_b\to \Lambda\pi^+\pi^-$ and $\Lambda_b\to p\bar K^0 \pi^-$ 
would  be regarded as
the first evidences of the scalar meson productions in the anti-triplet $b$-baryon decays.
The smaller predicted value of ${\cal B}(\Lambda_b\to \Lambda K^+K^-)$ indicates the existences of other 
resonant contributions to the decay, such as $\Lambda_b\to K^-(N^{*+}\to)\Lambda K^+$.
\end{abstract}

\maketitle
\section{introduction}
The LHCb Collaboration has recently measured the three-body $\Lambda_b$ decays~\cite{Aaij:2014lpa,LbtoLKK}, given by
\begin{eqnarray}\label{data1}
{\cal B}(\Lambda_b\to\Lambda \pi^+ \pi^-)&=&(4.6\pm 1.2\pm 1.4\pm 0.6)\times 10^{-6}\,,\nonumber\\
{\cal B}(\Lambda_b\to\Lambda K^+ K^-)&=&(15.9\pm 1.2\pm 1.2\pm 2.0)\times 10^{-6}\,,\nonumber\\
{\cal B}(\Lambda_b\to p \bar K^0 \pi^-)&=&(12.6\pm 1.9\pm 0.9\pm 3.4\pm 0.5)\times 10^{-6}\,.
\end{eqnarray}
However,  the present available calculations in the literature show that
${\cal B}(\Lambda_b\to \Lambda \rho^0\,,\rho^0\to \pi^+\pi^-)$ is
merely in the range of $10^{-9}-10^{-7}$~\cite{Guo:1998eg,Arunagiri:2003gu,Leitner:2006nb},
which is much smaller than that 
in Eq.~(\ref{data1}).
Similarly, according to 
the measured ${\cal B}(\Lambda_b\to\Lambda\phi)=
(5.18\pm 1.04\pm 0.35^{+0.67}_{-0.62})\times 10^{-6}$~\cite{LbtoLphi} and 
the predicted ${\cal B}(\Lambda_b\to p K^{*-})
=(2.5\pm 0.3\pm 0.2\pm 0.3)\times 10^{-6}$~\cite{Hsiao:2014mua},
the resonant vector $\phi\to K^+ K^-$ and $K^{*-}\to \bar K^0\pi^-$ contributions
lead to ${\cal B}(\Lambda_b\to \Lambda K^+K^-,p \bar K^0\pi^-)=
(2.5\pm 0.6,1.7\pm 0.3)\times 10^{-6}$, which are also unable to explain the data in Eq.~(\ref{data1}). 
Clearly, there must be some undiscovered contributions 
to these three-body $\Lambda_b$ decays,  respectively.

In this study, we propose to use the resonant scalar mesons as the dominant productions to resolve the deficits for
the three-body decays in Eq.~(\ref{data1}).  
Explicitly,  we consider
the scalar meson decays of $f_0(980,1500)\to (\pi^+\pi^-,\,K^+ K^-)$ 
and  $K^{*-}_0(1430)\to \bar K^0\pi^-$ through the charmless two-body  processes of
$\Lambda_b\to\Lambda f_0(980,1500)$ and $\Lambda_b\to p K^{*-}_0(1430)$
to produce 
the three-body $\Lambda_b$ decays in Eq.~(\ref{data1}).
We will demonstrate that 
${\cal B}(\Lambda_b\to\Lambda \pi^+ \pi^-,\,\Lambda K^+ K^-)$ and 
${\cal B}(\Lambda_b\to p \bar K^0 \pi^-)$ can be taken as
the first evidences for the scalar meson productions 
in the charmless two-body $\Lambda_b$ decays. Our present study will be useful
to distinguish the resonant contributions of the two-quark, tetraquark and glueball bound states, respectively,
similar to the tetraquark and scalar meson searches 
in the charmful two-body cases~\cite{Sharma:2011zzb,XYZ}.

Note that theoretical studies are still controversial concerning 
the underlying structures of the scalar 
mesons~\cite{pdg,Close:2002zu}.
For example, $f_0(980)$ is one of the scalar mesons lighter than 1 GeV 
to be identified as either the two-quark or four-quark (tetraquark)
bound states~\cite{Weinberg:2013cfa,Aaij:2014siy},
while $f_0(1500)$ and $K^{*-}_0(1430)$ heavier than 1 GeV
belong to the conventional $q\bar q$ nonet but with $f_0(1500)$ identified to 
primarily consist of either the glueball or the $s\bar s$ bound states~\cite{Hsiao:2014dva}.
Nevertheless, the scalar quark currents in the decaying processes favor 
the formations of the scalar mesons due to the quantum numbers of $J^{PC}=0^{++}$,
causing the enhanced branching ratios compared to the decays with the 
recoiled vector mesons of $\rho^0$, $\phi$, and $K^{*-}$.
\section{Formalism}
\begin{figure}[t!]
\centering
\includegraphics[width=2.2in]{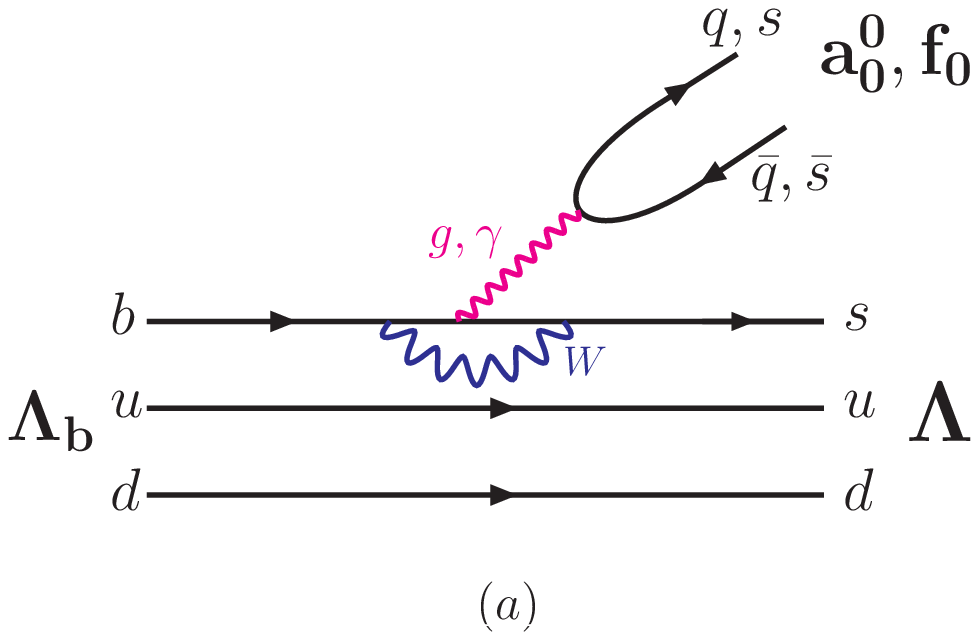}
\includegraphics[width=2.2in]{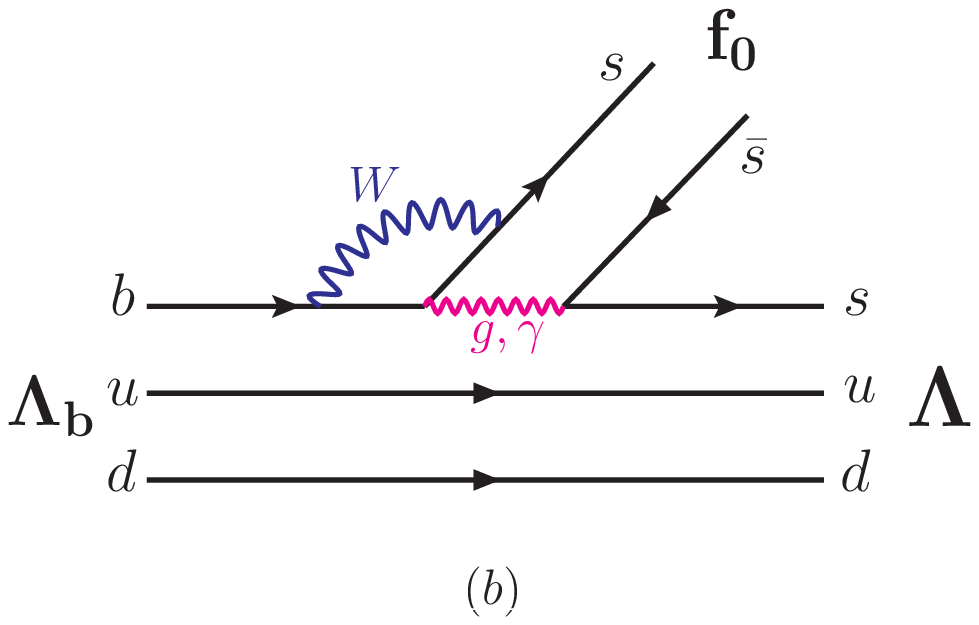}
\includegraphics[width=2in]{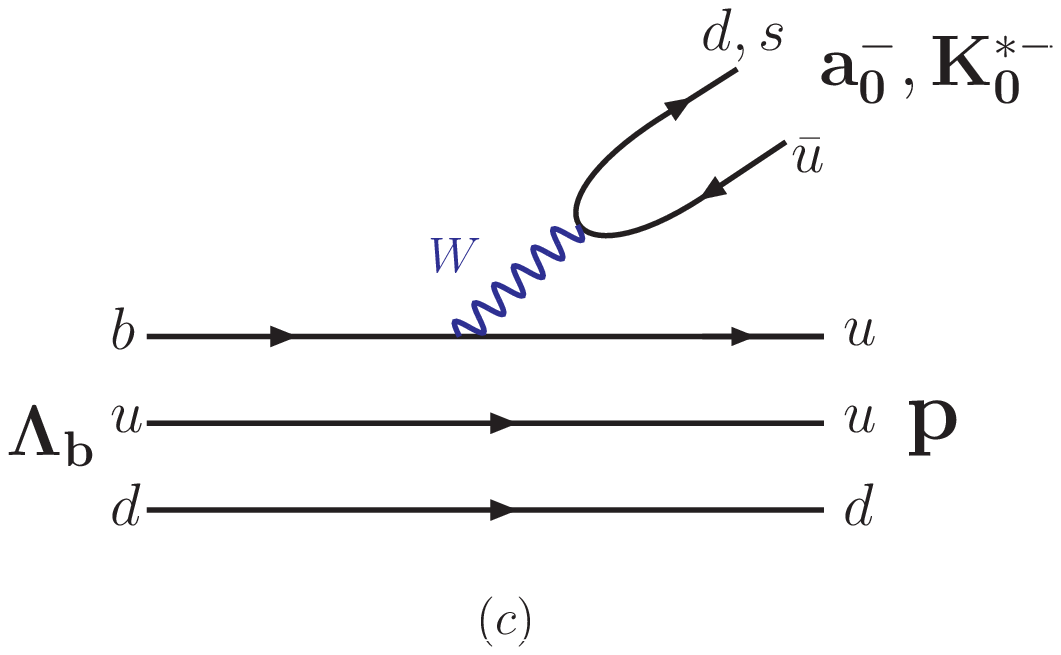}
\includegraphics[width=2in]{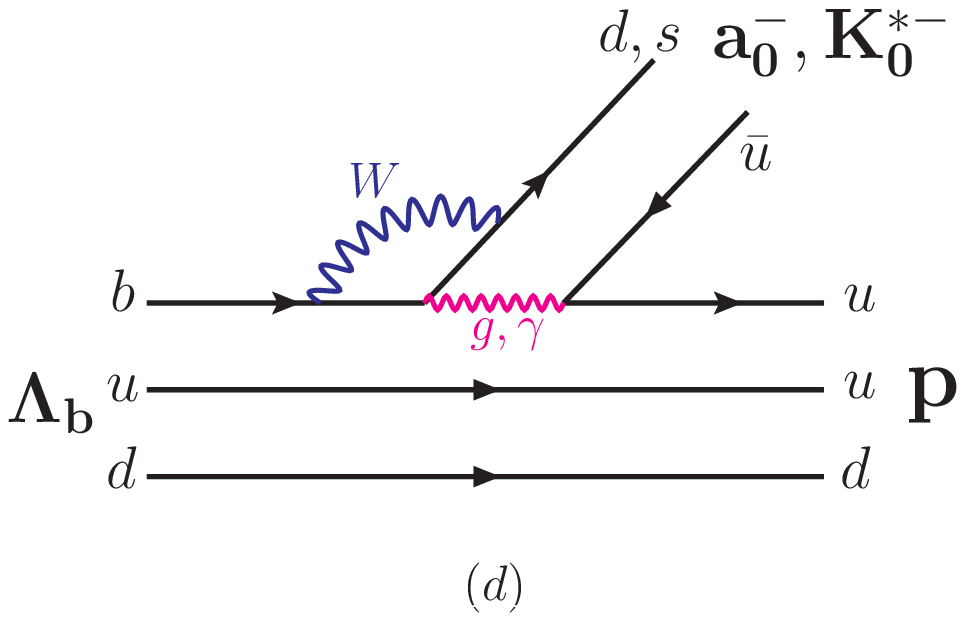}
\includegraphics[width=2in]{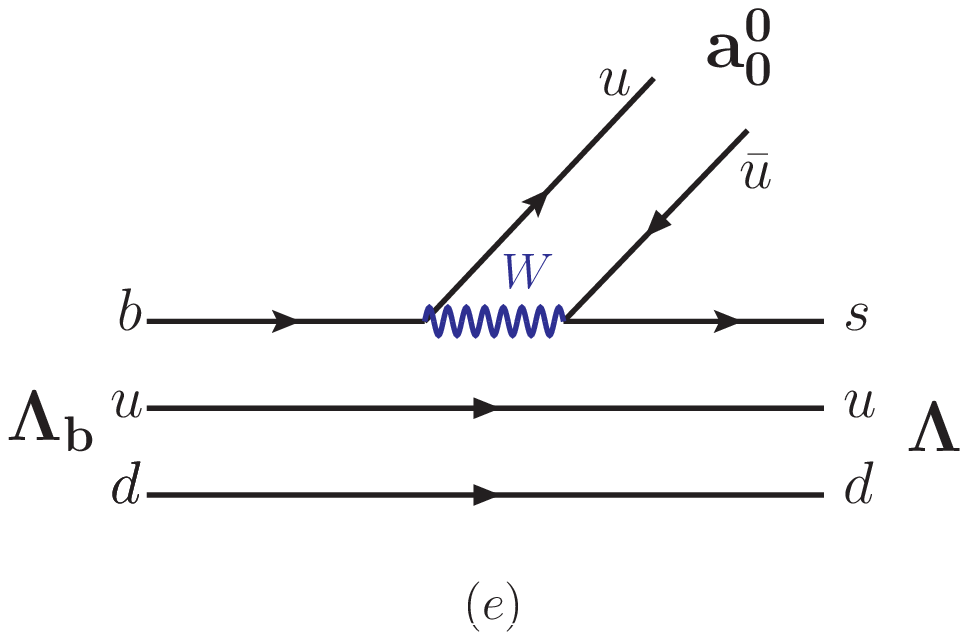}
\caption{Diagrams for the $\Lambda_b^0\to{\cal B}_n S$ decays,
where $S$ denotes as the scalar mesons, 
such as $a_0^0$, $f_0$, $K^{*-}_0$, and $a_0^-$.}\label{dia}
\end{figure}
In accordance with the diagrams depicted in Fig.~\ref{dia}, the amplitudes of 
$\Lambda_b\to {\cal B}_n S$ 
via the effective Hamiltonian can be decomposed as 
the matrix elements of the $\Lambda_b\to {\cal B}_n$ baryon transitions as well as
the vacuum to scalar meson productions ($0\to S$), given by~\cite{ali} 
\begin{eqnarray}\label{amp1a}
{\cal A}(\Lambda_b\to \Lambda f_0)&=&\frac{G_F}{\sqrt 2}
\bigg\{\alpha_3\langle f_0|\bar s\gamma_\mu s|0\rangle\langle 
\Lambda|\bar s\gamma_\mu(1-\gamma_5)b|\Lambda_b\rangle
+\alpha_6^s\langle f_0|\bar s s|0\rangle \langle \Lambda|\bar s(1-\gamma_5)b|\Lambda_b\rangle\bigg\}\,,
\nonumber\\
{\cal A}(\Lambda_b\to p S_0^-)&=&\frac{G_F}{\sqrt 2}
\bigg\{\alpha_1^q\langle S_0^-|\bar q\gamma_\mu u|0\rangle\langle 
p|\bar u\gamma_\mu(1-\gamma_5)b|\Lambda_b\rangle
+\alpha_6^q\langle S_0^-|\bar q u|0\rangle \langle p|\bar u(1-\gamma_5)b|\Lambda_b\rangle\bigg\}\,,\nonumber\\
{\cal A}(\Lambda_b\to \Lambda a_0^0)&=&\frac{G_F}{\sqrt 2}
\alpha_2\langle a_0^0|\bar q\gamma_\mu q|0\rangle\langle 
\Lambda|\bar s\gamma_\mu(1-\gamma_5)b|\Lambda_b\rangle\,,
\end{eqnarray}
where $G_F$ is the Fermi constant, $f_0= f_0(980,1500)$, 
$S_0^-=(K^{*-}_0(800,1430),a_0^-(980,1450))$ for $q=(s,d)$, 
and $a_0^0=a_0^0(980,1450)$ for $q=u$ or $d$, while
the parameters $\alpha_i$ are given by
\begin{eqnarray}\label{alpha1}
\alpha_1^q&=&V_{ub}V_{uq}^*a_1-V_{tb}V_{tq}^* a_4^q,\nonumber\\
\alpha_2&=&V_{ub}V_{us}^*\,a_2-V_{tb}V_{ts}^* 3a_9/2,\nonumber\\
\alpha_3&=&-V_{tb}V_{ts}^*(a_3+a_4^s+a_5-a_9/2),\nonumber\\
\alpha_6^q&=&V_{tb}V_{tq}^*\,2a_6^q,
\end{eqnarray}
with $V_{q_1q_2}$ the Cabibbo-Kobayashi-Maskawa (CKM) matrix elements, 
where $a_i\equiv c^{eff}_i+c^{eff}_{i\pm1}/N_c$ for $i=$odd (even)
consist of the effective Wilson coefficients $c_i^{eff}$ defined in Ref.~\cite{ali}
with the color number $N_c$ 
 fixed to be 3 in the naive factorization. Nonetheless,
in the generalized factorization~\cite{ali} used in this study,
one is allowed to float $N_c$ from 2 to $\infty$ to estimate the non-factorizable effects,
which are taken as a part of the theoretical uncertainty.
Note that although the use of the factorization method has been successful in various baryonic decays~\cite{Hsiao:2016amt},
the present case with scalar mesons, such as $f_0$, may contain some uncontrollable uncertainty due to
the less-known meson structures.

The matrix elements of the $0\to S$ production are given by~\cite{Wang:2006ria,Cheng:2005nb}
\begin{eqnarray}\label{ffs1}
\langle S|\bar q_2 \gamma_\mu q_1|0\rangle=f_S q_\mu\,,
\langle S|\bar q_2 q_1|0\rangle=m_S \bar f_S\,,
\end{eqnarray}
with $f_S$ and $\bar f_S$ the decay constants, where
$q_\mu$ is the four momentum vector. 
For the neutral scalar mesons, one has $f_{f_0}=f_{a^0_0}=0$,
 due to the charge conjugation invariance as well as the conservation of the vector current,
such that the $\alpha_{2,3}$ terms in Eq.~(\ref{amp1a}) vanish, resulting in
${\cal B}(\Lambda_b\to \Lambda a_0^0(980,1450))=0$.
For the charged ones, 
$f_S$ and $\bar f_S$ are related as $m_S f_S=(m_{q_2}-m_{q_1})\bar f_S$
by using equation of motion,
such that $f_{a_0^-}=\bar f_{a_0^-}(m_d-m_u)/m_{f_{a_0^-}}$ causes 
the suppressed $\alpha_1^d$ term that contributes to $\Lambda_b\to p a_0^-(980,1450)$ in Eq.~(\ref{amp1a}).
The matrix elements of the $\Lambda_b\to {\cal B}_n$ baryon transitions
are parameterized as~\cite{Hsiao:2014mua}
\begin{eqnarray}\label{ffs2}
\langle {\cal B}_n|\bar q \gamma_\mu b|\Lambda_b\rangle&=&
\bar u_{{\cal B}_n}[f_1\gamma_\mu+\frac{f_2}{m_{\Lambda_b}}i\sigma_{\mu\nu}q^\nu+
\frac{f_3}{m_{\Lambda_b}}q_\mu] u_{\Lambda_b}\,,\nonumber\\
\langle {\cal B}_n|\bar q \gamma_\mu\gamma_5 b|\Lambda_b\rangle&=&
\bar u_{{\cal B}_n}[g_1\gamma_\mu+\frac{g_2}{m_{\Lambda_b}}i\sigma_{\mu\nu}q^\nu+
\frac{g_3}{m_{\Lambda_b}}q_\mu]\gamma_5 u_{\Lambda_b}\,,\nonumber\\
%
%
\langle {\cal B}_n|\bar q(1-\gamma_5)b|\Lambda_b\rangle&=&
\bar u_{{\cal B}_n}(g_S\gamma_\mu-g_P\gamma_\mu \gamma_5)u_{\Lambda_b}\,,
\end{eqnarray}
where $f_{i}(g_i)$ (i=1,2 and 3) and $g_{S(P)}$ are the form factors, 
where $f_1=g_1$ and $f_{2,3}=g_{2,3}=0$ are derived by
the $SU(3)$ flavor and $SU(2)$ spin symmetries~\cite{Brodsky1,Hsiao:2015cda},
 in agreement with QCD models~\cite{CF,Wei:2009np,Gutsche:2013oea}.
From equation of motion, one obtains that $g_{S}=a_{S}f_1$ and $g_{P}=a_{P}g_1$ 
with $a_{s,p}=(m_{\Lambda_b}\mp m_{{\cal B}_n})/(m_b\mp m_q)$.

\section{Numerical Results and Discussions }
For our numerical analysis, 
the CKM matrix elements in the Wolfenstein parameterization 
are presented as
\begin{eqnarray}
&&(V_{ub},\,V_{tb})=(A\lambda^3(\rho-i\eta),1)\,,\nonumber\\
&&(V_{ud},\,V_{td})=(1-\lambda^2/2,A\lambda^3)\,,\nonumber\\
&&(V_{us},\,V_{ts})=(\lambda,-A\lambda^2),
\end{eqnarray}
with $(\lambda,\,A,\,\rho,\,\eta)=(0.225,\,0.814,\,0.120\pm 0.022,\,0.362\pm 0.013)$~\cite{pdg}.
To estimate the non-factorizable effects
in the generalized factorization approach~\cite{ali}, $a_i$
are taken as floating numbers for $N_c$ from 2 to $\infty$. 
The specific values of
$a_i$ with $N_c=2$, 3 and $\infty$ are given in Table~\ref{alpha_i}.
{\color{blue}
\begin{table}[t]
\caption{The parameters $a_i$ with $N_c=2,\,3$, and $\infty$
to estimate the non-factorizable effects in the generalized factorization.}\label{alpha_i}
\begin{tabular}{|c|ccc|}
\hline
$a_i$ & $N_c=2$ &  $N_c=3$ & $N_c=\infty$ \\\hline
$a_1$ & $0.98$ &  $1.05$ & $1.17$ \\
$10^{4}a_3$ & $-13.1-15.6i$ &  $72.4$ & $243.2+31.2i$ \\
$10^{4}a_4^d$ & $-377.6-34.7i$ &  $-417.2-37.0i$ & $-496.5-41.6i$ \\
$10^{4}a_4^s$ & $-391.0-77.9i$ &  $-431.6-83.1i$ & $-512.6-93.5i$ \\
$10^{4}a_5$ & $-174.1-15.6i$ &  $-65.8$ & $150.7+31.2i$ \\
$10^{4}a_6^d$ & $-560.7-34.7i$ &  $-584.9-37.0i$ & $-633.4-41.6i$ \\
$10^{4}a_6^s$ & $-574.1-77.9i$ &  $-599.3-83.1i$ & $-649.5-93.5i$ \\
$10^{4}a_9$ & $-93.5-2.2i$ &  $-99.8-2.2i$ & $-112.3-2.2i$ \\
\hline
\end{tabular}
\end{table}
}
With the double-pole momentum dependences, we have
$f_1(g_1)={C_{{\cal B}_n}}/{(1-q^2/m_{\Lambda_b}^2)^2}$
with 
$C_p=\sqrt {3/2}C_\Lambda=0.136\pm 0.009$~\cite{Hsiao:2014mua,Hsiao:2015cda}.
The decay constants are scale ($\mu$)-dependent, adopted to be~\cite{Cheng:2005nb}
\begin{eqnarray}
(\bar f_{f_0(980)}, \bar f_{f_0(1500)})&=&
(460\pm 25,605\pm 60)\,\text{MeV}\,,\nonumber\\
(\bar f_{K^*_0(800)},\bar f_{K^*_0(1430)})&=&
(420\pm 25,550\pm 60)\,\text{MeV}\,,\nonumber\\
(\bar f_{a_0^-(980)},\bar f_{a_0^-(1450)})&=&
(450\pm 25,570\pm 60)\,\text{MeV}\,,
\end{eqnarray}
with $\mu=2.1$ GeV, where the  model with the $q\bar q$ states for the scalar mesons
has been assumed. 
The branching ratios for the scalar meson productions in the two-body $\Lambda_b$ decays are shown in Table~\ref{tab2}.

\begin{table}[t]
\caption{Our numerical results for the branching ratios of  the two-body decays in unit of $10^{-6}$, 
where the first, second and third errors are from the non-factorizable effects,
 form factors and decay constants, respectively.}\label{tab2}
\begin{tabular}{|c|c|}
\hline
decay modes  &  our results \\\hline
$\Lambda_b\to \Lambda f_0(980)$
&$2.9^{+0.5}_{-0.2}\pm 0.4\pm 0.3$\\
$\Lambda_b\to \Lambda f_0(1500)$                     
&$12.4^{+2.2}_{-1.0}\pm 1.7\pm 2.6$ \\
$\Lambda_b\to p K^{*-}_0(800)$       
&$1.9^{+0.4}_{-0.2}\pm 0.3\pm 0.2$ \\
$\Lambda_b\to p K^{*-}_0(1430)$       
&$14.1^{+2.5}_{-1.2}\pm 1.9\pm 3.2$ \\
$\Lambda_b\to p a^-_0(980)$       
&$(7.8^{+1.3}_{-0.6}\pm 1.0\pm 0.9)\times 10^{-2}$ \\
$\Lambda_b\to p a^-_0(1450)$       
&$(6.3^{+1.1}_{-0.5}\pm 0.9\pm 0.1)\times 10^{-1}$ \\
$\Lambda_b\to \Lambda a_0^0(980,1450)$       
&0 \\
\hline
\end{tabular}
\end{table}
In Table~\ref{tab2}, 
the three errors correspond to
the uncertainties  from the non-facotrizable effects in $\alpha_i$ with $N_c=2-\infty$ illustrated in Table~\ref{alpha_i},
 form factors, and  decay constants, respectively. With the combined errors,
we see that  
${\cal B}(\Lambda_b\to \Lambda f_0(980,1500))=(2.9\pm 0.7,12.4\pm 3.8)\times 10^{-6}$
and ${\cal B}(\Lambda_b\to p K^{*-}_0(800,1430))=(1.7\pm 0.5,14.1\pm 4.5)\times 10^{-6}$, 
which are mainly from   
the $\alpha_6^s$ terms due to the penguin contributions,
where the $\bar s u$ and $\bar s s$ scalar currents favor the formations of 
$f_0(980,1500)$ and $K^{*-}_0(800,1430)$, respectively. 
On the contrary,
since the vector currents disfavor those of $a_0^-(980,1450)$,
which are in accordance with the decay constants of 
$f_{a^-_0(980,1500)}\simeq (1.8,1.1)$ MeV, 
we obtain ${\cal B}(\Lambda_b\to p a^-_0(980,1450))$ in the range of $10^{-8}-10^{-7}$.    
Despite of the predictions of ${\cal B}(\Lambda_b\to \Lambda a_0^0(980,1450))=0$,
which are due to $f_{a^0_0(980,1500)}=0$, 
the picture of the tetraquark state, that is, $a_0^0(980)\equiv 1/\sqrt 2(u\bar u-d\bar d)s\bar s$,
may let the decays receive the $s\bar s$ scalar current,
such that whether or not the branching ratios of $\Lambda_b\to \Lambda a_0^0(980,1450)$ 
are equal to zero can be clear measurements
to test the underlying structures of the scalar mesons. 

For the three-body $\Lambda_b$ decays, 
with the two-body decay branching ratios in Table~\ref{tab2}, 
 connected to the resonant scalar meson data of
${\cal B}(f_0(980)\to \pi^+\pi^-,K^+ K^-)=(46\pm 6,16.1\pm 7.2)\%$~\cite{f0980topipi},
${\cal B}(f_0(1500)\to \pi^+\pi^-, K^+ K^-)=(23.3\pm 1.5,4.3\pm 0.5)\%$~\cite{pdg}, and
${\cal B}(K^{*-}_0(1430)\to \bar K^0 \pi^-)=(62.0\pm 6.6)\%$~\cite{pdg}, 
we get our results for the three-body $\Lambda_b$ decays shown 
in Table~\ref{tab3}.
\begin{table}[b]
\caption{Our numerical results for the  branching ratios of  the three-body $\Lambda_b$ decays in unit of $10^{-6}$.} 
\label{tab3}
\begin{tabular}{|c|c|c|}
\hline
decay modes  &  our results & data~\cite{Aaij:2014lpa,LbtoLKK} \\\hline
$\Lambda_b\to \Lambda\pi^+\pi^-$
&$4.2\pm 1.0$ & $4.6\pm 1.9$\\
$\Lambda_b\to \Lambda K^+ K^-$
&$3.5\pm 0.7$& $15.9\pm 2.6$\\
$\Lambda_b\to p\bar K^0 \pi^-$ 
&$10.4\pm 2.9$& $12.6\pm4.0$\\  
\hline
\end{tabular}
\end{table}
It is interesting to see that our result for $\Lambda_b\to \Lambda\pi^+\pi^-$
fits the data very well,
which would be 
 regarded as the first evidence for the scalar meson productions
in the charmless two-body $\Lambda_b$ decays. 
As a result, it is expected to find the resonant peaks for $f_0(980)$ and $f_0(1500)$
in the $\pi\pi$ invariant mass spectrum.
Like $\Lambda_b\to \Lambda\pi^+\pi^-$, our result of
${\cal B}(\Lambda_b\to p\bar K^0 \pi^-)=(10.4\pm 2.9)\times 10^{-6}$
not only alleviates the theoretical deficit but also explains
the data of $(12.6\pm 4.0)\times 10^{-6}$.
For $\Lambda_b\to \Lambda K^+ K^-$,
 our prediction due to the resonant scalar mesons
seems  much lower than that of the data.
However, it still helps us to mimic 
 the theoretical shortage,
and leaves room for other contributions,
such as the resonant $\Lambda_b\to K^-(N^{*+}\to)\Lambda K^+$ decay.
Since there exists the possible sign revealed in the $m^2(\Lambda K^+)$ 
from the Dalitz plot~\cite{LbtoLKK}, according to the study of the measured
${\cal B}(\Lambda_b\to J/\psi(N^*\to)pK^+)
\simeq (3.04\pm 0.55)\times 10^{-4}$~\cite{Aaij:2015tga,Hsiao:2015nna}, 
we estimate that ${\cal B}(\Lambda_b\to K^-(N^{*+}\to)\Lambda K^+)\simeq 10\times 10^{-6}$,
which can explain the deviation.

\section{Conclusions}
In sum, we have studied the charmless two-body $\Lambda_b$ decays
with the scalar mesons as final states. 
We predicted the first scalar meson productions in the $\Lambda_b$ decays,
such as ${\cal B}(\Lambda_b\to \Lambda f_0(980,1500))=
(2.9\pm 0.7,12.4\pm 3.8)\times 10^{-6}$ and 
${\cal B}(\Lambda_b\to p K^{*-}_0(800,1430))=(1.9\pm 0.5,14.1\pm 4.5)\times 10^{-6}$.
With the resonant $f_0(980,1500)\to \pi^+\pi^-$, we have obtained
${\cal B}(\Lambda_b\to \Lambda\pi^+\pi^-)=(4.2\pm 1.0)\times 10^{-6}$,
which can explain the data of $(4.6\pm 1.9)\times 10^{-6}$
much underestimated by the previous studies.
Similarly, we have shown that the resonant scalar meson contributions from 
$f_0(980,1500)$ and $K^*_0(1430)^-$
lead to 
${\cal B}(\Lambda_b\to \Lambda K^+ K^-)=(3.5\pm 0.7)\times 10^{-6}$ and 
${\cal B}(\Lambda_b\to p\bar K^0 \pi^-)=(10.4\pm 2.9)\times 10^{-6}$,
which alleviate the theoretical shortages compared to the current  observations, respectively.

\section*{ACKNOWLEDGMENTS}
The authors would like to thank Dr. Eduardo Rodrigues for useful discussions.
The work was supported in part by 
National Center for Theoretical Sciences, 
National Science Council (NSC-101-2112-M-007-006-MY3), and
MoST (MoST-104-2112-M-007-003-MY3).

\end{document}